\documentclass{aa}

\usepackage{txfonts}
\usepackage{graphicx}

\begin{document}

\title{Optics for X-ray telescopes: analytical treatment of the off-axis effective area of mirrors in optical modules}

\author{D. Spiga}

\institute{INAF/Osservatorio Astronomico di Brera, Via E. Bianchi 46, I-23807, Merate (LC) - Italy\\ \email{daniele.spiga@brera.inaf.it}}

\date{Received 3 January 2011 / Accepted 17 January 2011}

\abstract 
{Optical modules for X-ray telescopes comprise several double-reflection mirrors operating in grazing incidence. The concentration power of an optical module, which determines primarily the telescope's sensitivity, is in general expressed by its on-axis effective area as a function of the X-ray energy. Nevertheless, the effective area of X-ray mirrors in general decreases as the source moves off-axis, with a consequent loss of sensitivity. To make matters worse, the dense nesting of mirror shells in an optical module results in a mutual obstruction of their aperture when an astronomical source is off-axis, with a further effective area reduction.}
{To ensure the performance of X-ray optics for new X-ray telescopes (like NuSTAR, NHXM, ASTRO-H, IXO), their design entails a detailed computation of the effective area over all the telescope's field of view. While the effective area of an X-ray mirror is easy to predict on-axis, the same task becomes more difficult for a source off-axis. It is therefore important to develop an appropriate formalism to reliably compute the off-axis effective area of a Wolter-I mirror, including the effect of obstructions.}
{Most of collecting area simulation for X-ray optical modules has been so far performed along with numerical codes, involving ray-tracing routines, very effective but in general complex, difficult to handle, time consuming and affected by statistical errors. In contrast, in a previous paper we approached this problem from an analytical viewpoint, to the end of simplifying and speeding up the prediction of the off-axis effective area of unobstructed X-ray mirrors with any reflective coating, including multilayers.}
{In this work we extend the analytical results obtained: we show that the analytical formula for the off-axis effective area can be inverted, and we expose in detail a novel analytical treatment of mutual shell obstruction in densely nested mirror assemblies, which reduces the off-axis effective area computation to a simple integration. The results are in excellent agreement with the findings of a detailed ray-tracing routine.}
{}

\titlerunning{Optics for X-ray telescopes: analytical treatment of the off-axis effective area}
\authorrunning{D. Spiga}
\maketitle

\section{Introduction}
Optical modules for X-ray telescopes consist of a number of grazing incidence mirror shells with a common axis and focus. In a widespread design, the Wolter's, the mirrors comprise two consecutive segments, a paraboloid and a hyperboloid, in order to concentrate X-rays by means of a double reflection. The two reflections occur at the same incidence angle for X-rays coming from an on-axis source at astronomical distance (Van~Speybroeck~\& Chase~\cite{VanSpeybroeck}). The optical design of the module is primarily dependent on the required effective area, which determines the telescope's sensitivity.

The most representative indicator of the concentration power of an optical module is assumed in general to be the effective area on-axis. However, the effective area in general decreases as the X-ray source moves off-axis. This in turn diminishes the telescope's sensitivity, hence may represent a severe limitation to its {\it field of view}. For this reason, the prediction of the on- and off-axis effective area modules is a very important task in the development of new X-ray telescopes such as NuSTAR (Hailey~et al.~\cite{Hailey}), NHXM (Basso~et al.~\cite{Basso}), ASTRO-H (Kunieda~et al.~\cite{Kunieda}), and IXO (Bookbinder~\cite{Bookbinder}). 

While the effective area on-axis is in general easy to calculate for a typical Wolter-I mirror module configuration, the computation becomes more difficult off-axis, because of the variable incidence angles over the two reflecting surfaces and the variable fraction of singly-reflected X-rays that are not focused and contribute to the stray light (see e.g., Cusumano~et al.~\cite{Cusumano}). To make things worse, the assembled mirrors can shade each other if they are not spaced enough, which contributes to an even steeper decrease in the collecting area off-axis. A widespread method for performing the calculation, accounting for all these factors, has made use of accurate ray-tracing routines (see, e.g., Mangus~\& Underwood~\cite{Mangus}; Zhao~et~al.~\cite{Zhao}), which reconstruct the paths of a selection of X-rays impinging a mirror module. These numerical codes are in general accurate, because they simulate the real incidence of rays on the optical system. However, they are complex and time consuming, especially whenever the simulation includes wide-band multilayer coatings to extend the reflectivity beyond 10 keV (Joensen~et~al.~\cite{Joensen}; Tawara~et~al.~\cite{Tawara}). The reason is that the multilayer reflectivity computation, which has to be performed for every ray traced, is a complex procedure especially for wideband multilayers, which comprise many ($\sim$ 200) couples of layers.

For this reason, although the ray-tracing approach should not be disregarded, it is interesting to derive analytical formulae for the effective area off-axis. In attempting to achieve this, Van~Speybroeck~\& Chase (\cite{VanSpeybroeck}) discovered, by analyzing the results of ray-tracing simulations, that the geometric collecting area of a Wolter-I mirror decreases with the off-axis angle $\theta$ of the source, according to the formula
\begin{equation}
	A_{\infty}(\theta) = A_{\infty}(0)\,\left(1-\frac{2\theta}{3\alpha_0}\right),
\label{eq:SC_formula}
\end{equation}
where $A_{\infty}(0)$ is the on-axis geometric area and $\alpha_0$ is the incidence angle for an astronomical source on-axis.

More recently, a semi-analytical method for computing the off-axis effective area has been applied to solve the problem of optimizing multilayer recipes to the telescope's field of view (Mao~et~al.~\cite{Mao99}; Mao~et~al.~\cite{Mao00}; Madsen~et~al.~\cite{Madsen}). In this approach, the effective area is computed by integrating the mirror reflectivity over the incidence angles, after weighting it over an appropriate function, $W_{\mathrm{inc}}$, derived from a ray-tracing. 

In a previous paper (Spiga~et~al.~\cite{Spiga2009}), we already derived a {\it completely} analytical method for computing the off-axis effective area of a Wolter-I mirror with any reflective coating. In a subsequent article (Spiga~\&~Cotroneo~\cite{Spiga2010}), we developed this formalism, deriving the analytical expression of the aforementioned $W_{\mathrm{inc}}$ function, which represents the distribution of the off-axis effective area over the incidence angles of rays, and we also used it to face the problem of multilayer optimization (Cotroneo~et~al.~\cite{Cotroneo2010}). The results were accurately verified as well, by comparison with the outcomes of a ray-tracing program. However, in these previous works, we did not consider the mutual shading (also known as {\it vignetting}) of mirrors, which may occur in mirror modules when shells are densely nested together. While in general the mirror module is designed to avoid any vignetting on-axis, the problem may arise for sources off-axis and cause a further loss of effective area. Consequently, the results could be applied only to single mirror shells, or to mirror assemblies for which the mutual obstruction is known to be negligible over all the field of view.

In this work, we overcome these limitations and extend the developed formalism to the general case of obstructed Wolter-I mirrors in X-ray optical modules. We still assume that the mirror profile can be approximated by a double cone as far as the sole effective area is concerned, a condition in general fulfilled by optics with large $f$-numbers. In Sect.~\ref{single}, we briefly review the results obtained for unobstructed single mirror shells, and in addition we show how the analytical formalism can be inverted to derive the product of the two reflectivities from the desired effective area variation with the off-axis angle. In Sect.~\ref{Vign}, we describe the geometrical parameters driving the nested mirror obstructions, derive the expression of the vignetting coefficients, and obtain {\it an analytical formula for the off-axis effective area of an obstructed mirror} (Eq.~(\ref{eq:area_total_fin})). We then derive in Sect.~\ref{Application} some analytical expressions for the obstructed geometric area, and we apply the results to the geometrical optimization of the module. In Sect.~\ref{Validation}, we prove the validity of the analytical formulae by means of a ray-tracing routine. The results are briefly summarized in Sect.~\ref{Conclusions}.

\section{Analytical formulae for unobstructed Wolter-I mirrors}\label{single}

We consider a Wolter-I mirror shell (Fig.~\ref{fig:wolter1}) with focal length~$f$. We denote with $R_0$ the radius {\it of the reflective surface} at the intersection plane and with $\alpha_0$ the incidence angle at the intersection plane for an astronomical source on-axis. They are related by the well-known relation 
\begin{equation}
	R_0 = f \tan(4\alpha_0).
\label{eq:Rf}
\end{equation}
We hereafter assume that all incidence angles are shallow, therefore $\tan(4\alpha_0) \approx 4\alpha_0$. The optical axis of the mirror is oriented in the $z$-axis direction. We define $L_1$ and $L_2$ to be the lengths of the parabolic and hyperbolic segments (hereafter named {\it primary} and {\it secondary}), which are supposed to be undeformed and very smooth. We denote with $R_{\mathrm M}$ and $R_{\mathrm m}$ the mirror radii at the entrance and exit pupil, respectively. We define $r_{\lambda}(\alpha)$ to be the coating reflectivity for a generic incidence angle $\alpha$, at the X-ray wavelength $\lambda$, and assume that the geometric optics can be applied, i.e., that $\lambda$ is small enough to avoid aperture diffraction effects but large enough for the scattering due to surface roughness to be negligible (Raimondi~\&~Spiga~\cite{RaiSpi}). The source is assumed to be at a finite, although very large, distance $D \gg L_{1,2}$. Finally, $\theta \geq 0$ denotes the off-axis angle, i.e., the angle formed by the source direction with the optical axis. The $x$-axis of this reference frame is oriented such that the source lies in the $xz$ plane, at $x > 0$ and $z > 0$. 

\subsection{A review of previous results}\label{review}

We briefly review the results about the computation of the off-axis effective area of unobstructed Wolter-I mirror shells, derived in previous works (Spiga~et~al.~\cite{Spiga2009}; Spiga~\& Cotroneo~\cite{Spiga2010}).
	
\begin{figure}
	\resizebox{\hsize}{!}{\includegraphics{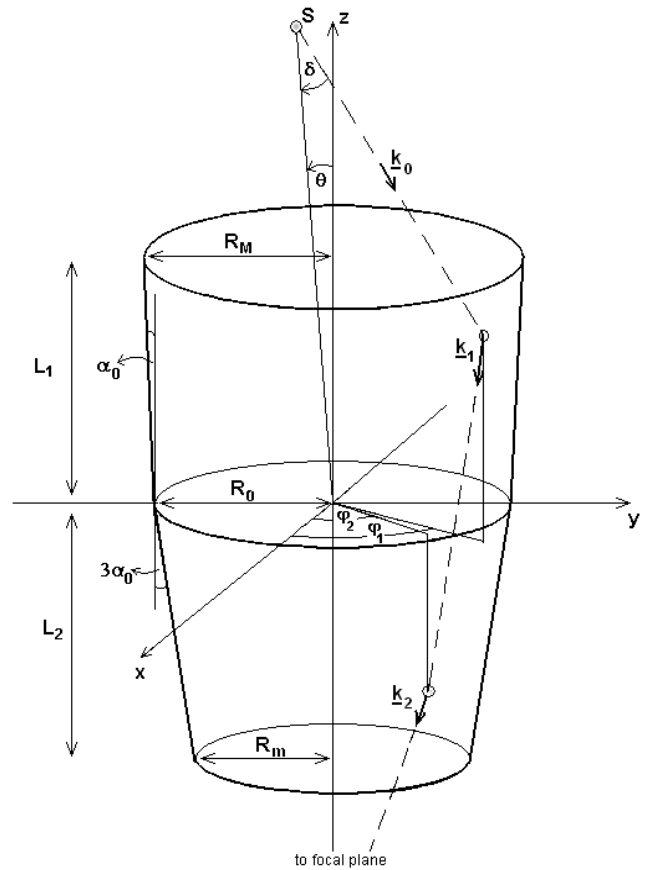}}
	\caption{Sketch of a grazing-incidence Wolter-I mirror shell with an off-axis X-ray source. We also show a ray direction vector before reflection~($\underline{k}_0$), after the first reflection ($\underline{k}_1$), and after two reflections ($\underline{k}_2$). }
	\label{fig:wolter1}
\end{figure}

\begin{enumerate}
\item{Owing to the shallow incidence angles, the double cone approximation is in general applicable to compute the effective area of Wolter-I mirrors. The error in the effective area estimation, by replacing a Wolter-I profile with a double cone, is on the order of $L/f$ or smaller, i.e., a few percent in real cases.}

\item{For a source off-axis, the incidence angles on the primary and secondary segment, $\alpha_1$ and $\alpha_2$, essentially depend on the polar angles $\varphi_1$ and $\varphi_2$ of the points where the ray is reflected (Fig.~\ref{fig:wolter1}). As long as $\theta$ is small, $\varphi_1 \approx \varphi_2$. We thus denote their nearly-common value with $\varphi$. If the polar angle $\varphi$ is measured from the $x$-axis, $\alpha_1$ and $\alpha_2$ have the expressions
	\begin{eqnarray}
		\alpha_1 &=& \alpha_0+\delta-\theta\cos\varphi \label{eq:angle1},\\
		\alpha_2 &=& \alpha_0-\delta+\theta\cos\varphi \label{eq:angle2},
	\end{eqnarray}
where $\delta = R_0/D$ is the X-ray semi-divergence due to the distance of the source: for an astronomical source, $\delta \simeq 0$. Equations~(\ref{eq:angle1}) and~(\ref{eq:angle2}) are valid if $\alpha_1\ge 0$, $\alpha_2 \ge 0$. We note that
\begin{equation}
	\alpha_1+\alpha_2=2\alpha_0.
	\label{eq:anglesum}
\end{equation}	}

\item{The geometric ratio of rays that undergo the double reflection to those impinging the primary segment is expressed by the {\it vignetting factor}
\begin{equation}
	V(\varphi) = \frac{L_2\alpha_2}{L_1\alpha_1},
	\label{eq:2reflvign}
\end{equation}
with the constraint that $0 < V(\varphi) <1$.}

\item{Using Eq.~(\ref{eq:2reflvign}), it is easy to derive an integral formula for the effective area of the mirror shell
	\begin{equation}
		A_D(\lambda, \theta) = 2R_0 \int_0^{\pi}\!\!(L\alpha)_{\mathrm{min}} \,r_{\lambda}(\alpha_1)\,r_{\lambda}(\alpha_2) \,\mbox{d}\varphi,
	\label{eq:Aeff_fin_offaxis}
	\end{equation}
	where
	\begin{equation}
		(L\alpha)_{\mathrm{min}} = \max\left\{0, \min[L_1 \alpha_1(\varphi), L_2 \alpha_2(\varphi)]\right\}.
		\label{eq:Lalpha_min}
	\end{equation}}

\item{In the frequent case $L_1 = L_2~(= L)$, Eq.~(\ref{eq:Aeff_fin_offaxis}) reduces to
	\begin{equation}
		A_{D}(\lambda, \theta) = 2R_0 L\int_0^{\pi}\!\!\alpha_{\mathrm{min}}\,r_{\lambda}(\alpha_1)\,r_{\lambda}(\alpha_2)\, \mbox{d}\varphi,
	\label{eq:Aeff_fin}
	\end{equation}
	where $\alpha_{\mathrm{min}} = \min[\alpha_1(\varphi), \alpha_2(\varphi)]$ if positive, and zero otherwise. The expression of the $\alpha_{\mathrm{min}}$ angle can be written in a compact form
	\begin{equation}
		\alpha_{\mathrm{min}} = \max(0,\,\alpha_0-|\delta-\theta\cos\varphi|),
	\label{eq:theta_min}
	\end{equation}}

\item{In the ideal case of a perfectly-reflecting mirror, i.e., setting $r =1$ for any $\alpha$ and $\lambda$, the integral in Eq.~(\ref{eq:Aeff_fin}) can be solved explicitly. For instance, if $\delta = 0$ and $\theta =0$, we obtain
	\begin{equation}
		A_{\infty}(0) =2\pi R_0 L \,\alpha_0,
	\label{eq:Ageom_1}
	\end{equation}
	whereas, if $\delta = 0$ and $0<\theta<\alpha_0$,
	\begin{equation}
		A_{\infty}(\theta) = A_{\infty}(0)\, \left(1- \frac{2\theta}{\pi\alpha_0}\right),
	\label{eq:Ageom_2}
	\end{equation}
	where -- and heretofore -- we drop the dependence of the $A$'s on $\lambda$ to denote the geometric mirror areas. Equation~(\ref{eq:Ageom_2}), after approximating $\pi\approx 3$, becomes Eq.~(\ref{eq:SC_formula}), the expression found by Van Speybroeck~\&~Chase in~\cite{VanSpeybroeck}. For $\theta > \alpha_0$, the geometric area has the more complicated expression 
	\begin{equation}
		A_{\infty}(\theta) = A_{\infty}(0)\, \left[1-\frac{2}{\pi}\left(\frac{\theta}{\alpha_0}-\sqrt{\frac{\theta^2}{\alpha_0^2}-1}+\arccos\frac{\alpha_0}{\theta}\right)\right].
	\label{eq:Ageom_3}
	\end{equation}}
	
\item{If $\delta >0$, more analytical expressions can be obtained by solving the integral in Eq.~(\ref{eq:Aeff_fin}). The predictions are in very good agreement with the findings of detailed ray-tracing routines. For more details, we refer to Spiga~et~al.~\cite{Spiga2009}.}

\item{If $\delta =0$ {\it and} $0 <\theta < \alpha_0$, we obtain an alternative form of Eq.~(\ref{eq:Aeff_fin}), by changing the integration variable from $\varphi$ to $\alpha_1$,
\begin{equation}
	A_{\infty}(\lambda, \theta) =4 R_0 L \int_{\alpha_0-\theta}^{\alpha_0}\!\frac{\alpha_1\,r_{\lambda}(\alpha_1)\,r_{\lambda}(\alpha_2)}{\sqrt{\theta^2-(\alpha_0-\alpha_1)^2}}\, \mbox{d}\alpha_1.
	\label{eq:Aeff_alpha}
\end{equation}
This equation can be used to derive the effective area as a function of $\theta$, at a fixed $\lambda$.}
\end{enumerate}

\subsection{Inverse computation: from $A_{\infty}(\lambda, \theta)$ to the mirror reflectivity}\label{invcomp}
\begin{figure}
	\resizebox{\hsize}{!}{\includegraphics{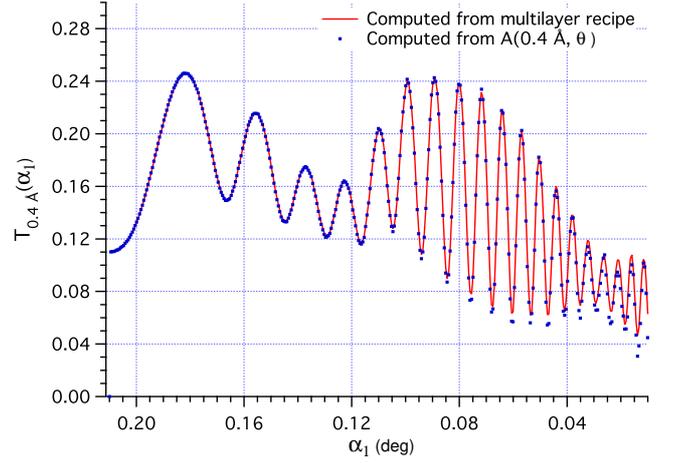}}
	\caption{The product of the reflectivities of the two segments of a Wolter-I shell with $\alpha_0$~=~0.212~deg, at 30 keV. The reflective surface is a multilayer coating, a Pt/C graded stack with 200 couples of layers. The d-spacing in the multilayer follows the power-law model $d(k) = a\,(b+k)^{-c}$ (Joensen~et~al.~\cite{Joensen}), with $a$~= 77.4~\AA, $b$~= -0.94, $c$~= 0.223, a Pt thickness ratio $\Gamma$= 0.42, and a 4~\AA~roughness rms. The solid line is directly computed from the multilayer structure, whilst the dots are the result of the inverse computation from the effective area in Fig.~\ref{fig:effarea}.}
	\label{fig:refl_prod}
\end{figure}
\begin{figure}
	\resizebox{\hsize}{!}{\includegraphics{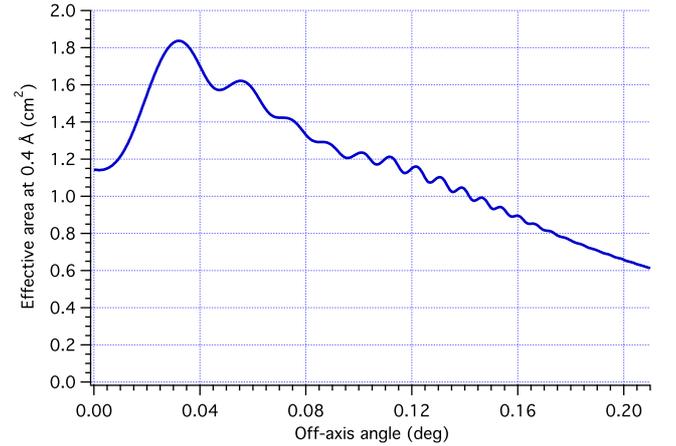}}
	\caption{The effective area at 30 keV ($\lambda$ = 0.4 \AA) of a Wolter-I mirror shell with $R_0$~= 148.5~mm, $L$~= 300~mm, $f$~= 10~m, $\alpha_0$~=~0.212~deg, and a reflectivity product as shown in Fig.~\ref{fig:refl_prod} (solid line). The effective area was used to re-derive the $T_{\lambda}(\alpha_1)$ function (Fig.~\ref{fig:refl_prod}, dots).}
	\label{fig:effarea}
\end{figure}

As a further development of the equations reported in the previous section for unobstructed mirrors, we hereafter provide the inverse formula of Eq.~(\ref{eq:Aeff_alpha}). For a given $\lambda$, from the effective area variation with $\theta$ in $[0, \alpha_0]$, for $\delta =0$, we derive the product of the reflectivities of the two segments
\begin{equation}
	T_{\lambda}(\alpha_1) \stackrel{\mathrm{def}}{=}r_{\lambda}(\alpha_1)\,r_{\lambda}(\alpha_2),
	\label{eq:prod_def}
\end{equation}
where $\alpha_2 = 2\alpha_0-\alpha_1$ (Eq.~(\ref{eq:anglesum})). Owing to the symmetry of $\alpha_1$ and $\alpha_2$ with respect to the $y$-axis when $\delta =0$, it is sufficient to compute $T_{\lambda}(\alpha_1)$ for $0<\alpha_1<\alpha_0$. To this end, we define $E_{\lambda}(\theta)$ to be the ratio of the effective area to the on-axis geometric area
\begin{equation}
	E_{\lambda}(\theta) =\frac{A_{\infty}(\lambda, \theta)}{A_{\infty}(0)},
	\label{eq:EA_norm}
\end{equation} 
and then it is easy to demonstrate (see Appendix~\ref{inversion}) that the $T_{\lambda}(\alpha_1)$ function, for $\alpha_1 \in (0, \alpha_0)$, can be computed from the integral equation 
\begin{equation}
	T_{\lambda}(\alpha_1) = \frac{\alpha_0}{\alpha_1}\int_{0}^{\pi/2}\!\!\sin t \left[\frac{\mbox{d}}{\mbox{d}\theta}(\theta \cdot E_{\lambda}(\theta))\right]_{\theta = \theta(t)}\!\!\!\!\!\!\!\mbox{d}t,
	\label{eq:refl_prod}
\end{equation}
where the expression in the [ ] brackets, after computing the derivative, is to be evaluated at 
\begin{equation}
	\theta(t)~=~(\alpha_0-\alpha_1)\sin t,
	\label{eq:tdef}
\end{equation}
where $0\leq t\leq \pi/2$ is a dummy integration variable. The condition $0<\alpha_1<\alpha_0$ then implies that $0<\theta<\alpha_0$: in other words, the $E_{\lambda}(\theta)$ function has to be known in the entire interval $(0, \alpha_0)$. By computing the integral in Eq.~(\ref{eq:refl_prod}) for {\it any} normalized effective area function $E_{\lambda}(\theta)$, one obtains the corresponding reflectivity product at $\lambda$ (Eq.~(\ref{eq:prod_def})), for $\alpha_1$ taking on values in the same interval. 

As a first example, we put in adimensional form the geometric vignetting for a source at infinity, Eq.~(\ref{eq:Ageom_2}),
\begin{equation}
	E(\theta) =1- \frac{2\theta}{\pi\alpha_0},
	\label{eq:refl_prod_geom0}
\end{equation}
and substitute this into Eq.~(\ref{eq:refl_prod}). We obtain
 \begin{equation}
	T_{\lambda}(\alpha_1) = \frac{\alpha_0}{\alpha_1}\int_{0}^{\pi/2}\!\left(\sin t - \frac{4(\alpha_0-\alpha_1)}{\pi\alpha_0}\sin^2t \right)\,\mbox{d}t,
	\label{eq:refl_prod_geom1}
\end{equation}
which can be immediately solved, yielding
 \begin{equation}
	T_{\lambda}(\alpha_1) = 1.
	\label{eq:refl_prod_geom2}
\end{equation}
This means that $r_{\lambda}(\alpha) =1$ for any incidence angle and wavelength, as expected. 

As a second example, we show the computation of a reflectivity product from the effective area of a mirror shell with a multilayer coating. The product of the two reflectivities at 30~keV, directly computed from the multilayer recipe, is shown in Fig.~\ref{fig:refl_prod}, as a function of $\alpha_1$. The effective area of the mirror shell at the same energy (Fig.~\ref{fig:effarea}) was computed in $(0, \alpha_0)$, using Eq.~(\ref{eq:Aeff_alpha}). Finally, we used Eq.~(\ref{eq:refl_prod}) to re-derive the $T_{\lambda}(\alpha_1)$ function from the effective area curve. The result of the inverse computation (Fig.~\ref{fig:refl_prod}, dots) closely matches the original reflectance product (Fig.~\ref{fig:refl_prod}, line).

\section{Obstructions in nested mirror modules}\label{Vign}
We now deal with a quantification of the obstructions that reduce the off-axis effective area of a mirror shell, when nested in a mirror module. Even if this effect is essentially a geometric shadowing, the effective area reduction depends on $\lambda$, because the incidence angles, and consequently the mirror reflectivity $r_{\lambda}(\alpha)$, vary over the mirror surface when the source is off-axis. This is especially true when multilayer coatings, which exhibit a complicated $r_{\lambda}(\alpha)$ function, are adopted. A reduction of the effective area is then relevant for the X-ray wavelengths that were reflected at the obstructed regions. 

The clear aperture of a mirror shell can be obstructed by several factors: the dense packing of shells, the structures for mechanical support of mirrors, and the presence of pre-collimators designed to reduce the stray light. However, we hereafter limit the discussion to the first point, i.e., the mutual obstruction of nested mirror shells, assuming that they are all co-axial, co-focal, and all have the same intersection plane at $z =0$ (Fig.~\ref{fig:obstruct}).

\subsection{Obstruction parameters}\label{obst_param}
\begin{figure}
	\resizebox{\hsize}{!}{\includegraphics{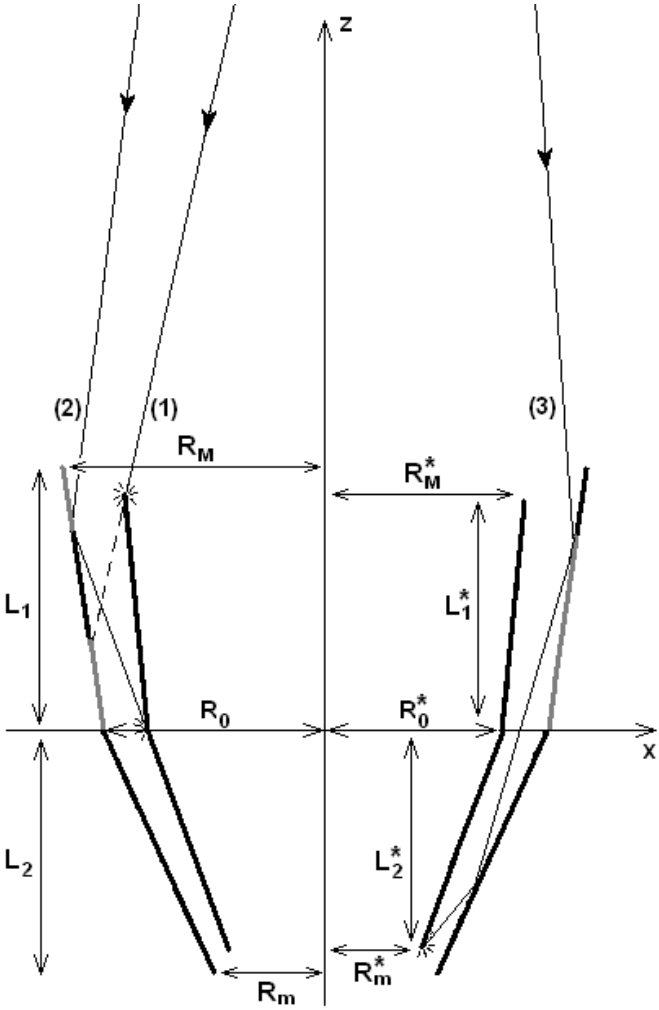}} 
	\caption{Obstruction in an assembly of mirror shells. Rays impinging a mirror shell can be blocked by the adjacent shell with smaller diameter in only three ways, as listed in the text. The impact points are highlighted. Other mirror shells are not shown. The obscured regions of the primary mirror are indicated in gray: angles and the mirror spacings are greatly exaggerated.}
	\label{fig:obstruct}
\end{figure}

We now draw our attention to a particular mirror shell in the mirror module (Fig.~\ref{fig:obstruct}), and adopt the same notation presented in Sect.~\ref{single}. The obstruction of this shell (which we refer to as a {\it reflective} shell) can be assumed to be caused solely by the shadow cast by the adjacent shell with smaller radius, which we refer to as {\it blocking} shell. The radius of its {\it outer} (i.e., non-reflective) surface at $z=0$ is denoted with $R_0^*$, which is forcedly smaller than $R_0$. We then define $R_{\mathrm M}^*$, $R_{\mathrm m}^*$, $L_1^*$, and $L_2^*$ to be respectively the maximum radius, the minimum radius, the primary segment length, and the secondary length of the {\it outer} surface of the blocking shell. We admit that, in general, $L_1 \neq L_1^*$ and $L_2 \neq L_2^*$. If $L_1^* = L_2^*$, we denote their common value with $L^*$. 

A simple geometrical construction shows (Fig.~\ref{fig:obstruct}) that an obstruction of the reflective shell can occur in three ways:
\begin{enumerate}
	\item{If rays intersect the blocking shell's primary segment before impinging the reflective shell;}
	\item{If, after the first reflection, rays impact onto the outer surface of the blocking shell's primary segment;}
	\item{If, after the second reflection, rays are blocked by the outer surface of the blocking shell's secondary segment.}
\end{enumerate}

The obstruction amount depends on the closeness of the two mirrors. This is often expressed along with $FF$, the {\it filling factor}
\begin{equation}
	FF = \frac{R_{\mathrm M}^*}{R_0}.
	\label{eq:ff}
\end{equation} 
The configuration with $FF=1$ causes each shell to exactly fit the clear area section of the adjacent shell, hence it maximizes the effective area on-axis at the expense of the off-axis area. If $FF>1$, the mirror assembly is self-obstructed even on-axis. For this reason and to enlarge the field of view of the optics, solutions with $FF<1$ are in general envisaged. 

\begin{figure}
	\resizebox{\hsize}{!}{\includegraphics{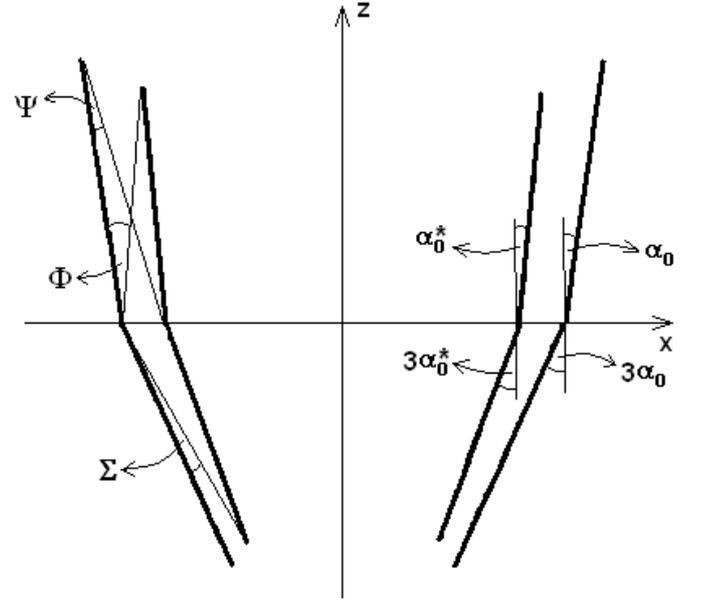}}
	\caption{Geometrical meaning of the obstruction parameters, $\Phi$, $\Psi$, and $\Sigma$, for a pair of Wolter-I nested shells.}
	\label{fig:obstruct_param}
\end{figure}

Nevertheless, it is convenient to adopt other parameters in the estimation of obstructions. The first of these is 
\begin{equation}
	\Phi = \frac{R_0-R_{\mathrm M}^*}{L_1^*}+\alpha_0,
	\label{eq:phi}
\end{equation}
which represents the maximum angle visible from the reflective shell at the intersection plane (see Fig.~\ref{fig:obstruct_param}), through the entrance pupil. Clearly, $\Phi > \alpha_0$ whenever $FF <1$, and $\Phi < \alpha_0$ if $FF >1$. As we later see, the $\Phi$ parameter drives the obstruction of the first kind, i.e., at the entrance pupil. 

The obstruction of the second kind, i.e., at the intersection plane, is chiefly determined by another parameter,
\begin{equation}
	\Psi = \frac{R_0-R_0^*}{L_1},
	\label{eq:psi}
\end{equation}
which denotes the clear angular aperture at the intersection plane, as seen from the maximum diameter of the reflective mirror shell (see Fig.~\ref{fig:obstruct_param}). The third obstruction parameter, which drives the third species of obstruction, is
\begin{equation}
	\Sigma = \frac{R_0-R_{\mathrm m}^*}{L_2^*}-3\alpha_0,
	\label{eq:sigma}
\end{equation}
which represents the angular aperture visible from the reflective shell at the intersection plane (see Fig.~\ref{fig:obstruct_param}), through the exit pupil. The importance of these angles becomes clearer in Sect.~\ref{obst_area}, when we derive the general formula for the obstructed mirror effective area (Eq.~(\ref{eq:area_total_fin})), which depends on $\Phi$, $\Psi$, and $\Sigma$ as parameters. 

The obstruction parameters can be related to each other
\begin{eqnarray}
	\Phi &=& \frac{L_1}{L_1^*}\Psi+(\alpha_0 -\alpha^*_0), \label{eq:rel1}\\
	 \Sigma &=& \frac{L_1}{L_2^*}\Psi -3(\alpha_0-\alpha^*_0), \label{eq:rel2}
\end{eqnarray}
where $\alpha_0^*$ is the on-axis incidence angle on the blocking shell. Using Eq.~(\ref{eq:Rf}), these relations can also be written as
\begin{eqnarray}
	\Phi &=& L_1\Psi\left(\frac{1}{L_1^*}+\frac{1}{4f}\right), \label{eq:rel1a}\\
	 \Sigma &=& L_1\Psi\left(\frac{1}{L_2^*} -\frac{3}{4f}\right). \label{eq:rel2a}
\end{eqnarray}
From Eqs.~(\ref{eq:rel1a}) and~(\ref{eq:rel2a}) it follows, in particular, that
\begin{itemize}
\item{if $L_1^* \leq L_1 \leq L_2^*$, then $\Sigma <\Psi <\Phi$;}
\item{if $L_1^* = L_1 = L_2^*$ and $f \gg L_1$, then $\Sigma \approx \Psi \approx \Phi$.}
\end{itemize}

\subsection{Vignetting coefficients}\label{obscoeff}
We now provide analytical expressions for the obstructions caused by the mirror nesting. To this end, we introduce a {\it vignetting coefficient}, $V_n(\varphi)$, with $n =1,2,3$, for every kind of obstruction as listed in Sect.~\ref{obst_param}. For an infinitesimal mirror sector located at the polar angle $\varphi$, the $n^{th}$ vignetting coefficient is defined as the fraction of primary segment's length left clear by the $n^{th}$ obstruction. This definition is analogous to that of $V(\varphi)$, the self-vignetting factor for double reflection (Eq.~(\ref{eq:2reflvign})), which we already treated in detail (Spiga~et~al.~\cite{Spiga2009}). We then have, in addition to $V(\varphi)$, three vignetting coefficients:
\begin{itemize}
	\item{$V_1(\varphi)$, due to the shadow cast by the blocking shell's primary segment before the first reflection;}
	\item{$V_2(\varphi)$, due to the shadow cast by the blocking shell's primary segment after the first reflection;}
	\item{$V_3(\varphi)$, due to the shadow cast by the blocking shell's secondary segment after the second reflection.}
\end{itemize} 
Firstly, we consider all vignetting factors to be independent of each other: the respective obstructed (i.e., lost) fractions of primary mirror length are denoted with $Q_n = 1-V_n$. 

As we anticipated in the previous section, it is the $\Phi$, $\Psi$, and $\Sigma$ parameters that primarily determine the vignetting coefficients. More precisely, they are functions of the incidence angles $\alpha_1$ and $\alpha_2$, and they can be computed via the following formulae
\begin{eqnarray}
	V_1(\alpha_1) &=& 1+ \frac{L_1^*(\Phi-\alpha_1)}{L_1\alpha_1},
	\label{eq:V1_}\\
	V_2(\alpha_1) &=& \frac{\Psi}{\alpha_1},
	\label{eq:V2_}\\
	V_3(\alpha_1) &=& 1+\frac{L_2^*(\Sigma-\alpha_2)}{L_1\alpha_1},
	\label{eq:V3_}
\end{eqnarray} 
where $\alpha_1+\alpha_2 = 2\alpha_0$, and with the usual constraint $0\leq V_n\leq~1$ for each $n$, otherwise we set $V_n$ to 0 or 1, respectively. The derivation of Eqs.~(\ref{eq:V1_}) to~(\ref{eq:V3_}) is not difficult, but quite lengthy, so it has been postponed to Appendix~\ref{deriV}. 

The dependence of the coefficients on $\varphi$ is obtained by substituting the expressions of $\alpha_1$ and $\alpha_2$ (Eqs.~(\ref{eq:angle1}) and~(\ref{eq:angle2})). We note that the obstructed region of the primary segment in the first and third kind of vignetting is located near the intersection plane, whereas $V_2$ results from an obscuration of the primary mirror near $z = +L_1$ (see Fig.~\ref{fig:obstruct}), in a precisely similar way to $V$. 

\begin{figure}
	\resizebox{\hsize}{!}{\includegraphics{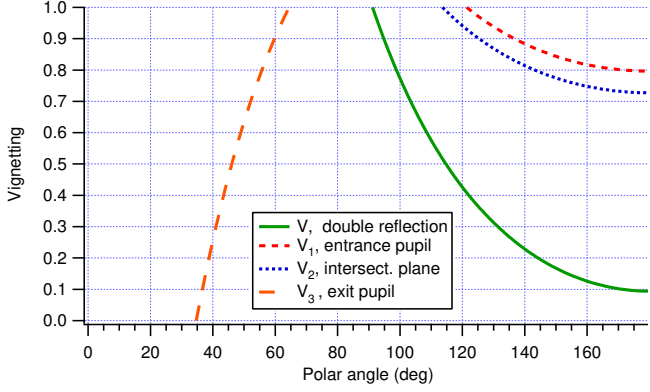}}
	\caption{Vignetting coefficients as a function of the polar angle for $L_2$~=~310~mm, $L_1$~= 300~mm, $L_2^*$~= 290~mm, $L_1^*$~= 280~mm, $\alpha_0$~=~0.3~deg, $\delta$~=~0~deg, $\theta$~=~0.25~deg, $\Phi$~=~0.43~deg, $\Psi$~=~0.4~deg, and $\Sigma$~=~0.408~deg. We also plot $V$, the vignetting factor for double reflection (Eq.~(\ref{eq:2reflvign})).}
	\label{fig:vign1}
\end{figure}

\begin{figure}
	\resizebox{\hsize}{!}{\includegraphics{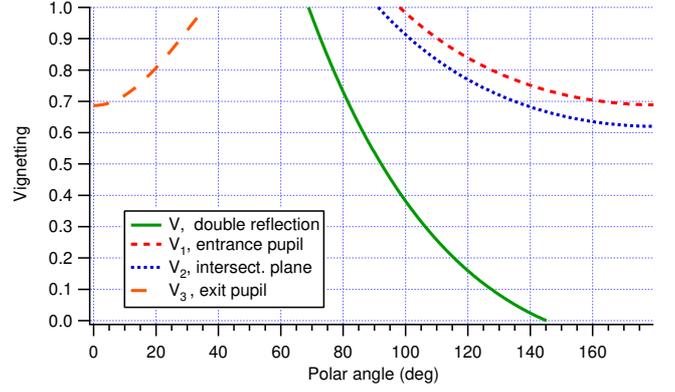}}
	\caption{Vignetting coefficients as a function of the polar angle for the same configuration as in Fig.~\ref{fig:vign1}, but this time the source is at a finite distance ($D$~=~127~m, $\delta$~=~0.095~deg). We note that the obstruction at the exit pupil is less severe than for a source at infinity.}
	\label{fig:vign2}
\end{figure}

Example plots of $V_1$, $V_2$, $V_3$, and $V$, as functions of $\varphi$, are drawn in Figs.~\ref{fig:vign1} and~\ref{fig:vign2}. The adopted values correspond to the case of two mirror shells with $R_0$~=~210~mm, $R_0^*$~=~207.9~mm, and $f$~=~10~m: the other parameter values are reported in the figure caption, including the lengths of the mirrors, which have been chosen to fulfill the relations $L_1^* < L_2^* < L_1 <L_2$. As we later see (Sect.~\ref{design}), this choice is not accidental: it represents a compromise to minimize all obstructions.

Equations~(\ref{eq:V1_}) to~(\ref{eq:V3_}) and Figs.~\ref{fig:vign1}, and~\ref{fig:vign2} show that:
\begin{itemize}
\item{If the conditions $\alpha_1 < \Phi$, $\alpha_1 < \Psi$, and $\alpha_2 <\Sigma$ are fulfilled at a given $\varphi$, there is no obstruction at that polar angle;} 
\item{If $\alpha_1 > \Psi$, but $V < V_2$ as in Figs.~\ref{fig:vign1} and~\ref{fig:vign2}, the obstruction at the intersection plane is ineffective, because all rays that were blocked would have missed the secondary segment;}
\item{The obstructions related to $V_1$ and $V_2$ are maximum at polar angles close to $\pi$;}
\item{The effect of $V_3$ is larger at $\varphi\approx 0$, where $\alpha_1$ is shallower and $\alpha_2$ is larger;}
\item{With the source at infinity, the obstruction related to $V_3$ can be very large, especially near $\varphi\approx 0$. In contrast, the effect of $V_3$ is mitigated if the source is at a finite distance, because $\alpha_2$ becomes smaller;} 
\item{If $L_1 \geq L_1^*$, then $V_1 > V_2$ for all $\varphi$;} 
\item{If $L_1 = L_1^*$ and $f \gg L_1$, then $\Phi \approx \Psi$ (Eq.~(\ref{eq:rel1a})) and also $V_1 \approx V_2$ for all $\varphi$.}
\end{itemize}

\subsection{General formula for the effective area of obstructed mirror shells}\label{obst_area}
We already pointed out that the obstructions described by $V$ and $V_2$ are located close to the maximum diameter of a Wolter-I mirror shell, whereas the ones related to $V_1$ and $V_3$ concern mainly the region close to its intersection plane. Therefore, $V$ and $V_2$ are in competition for the obstruction near $z = +L_1$, while $V_1$ and $V_3$ do the same near $z =0$ (Fig.~\ref{fig:obstruct}). The total obstruction at the generic polar angle $\varphi$ is then
\begin{equation}
	Q_{\mathrm{tot}}=\max(Q, Q_2, 0)+\max(Q_1, Q_3, 0) ,
	\label{eq:Q_total}
\end{equation} 
where the ``0'' value has been added to avoid negative obstructions. The corresponding total vignetting factor is $1-Q_{\mathrm{tot}}$, i.e., 
\begin{equation}
	V_{\mathrm{tot}}=\min(V, V_2, 1)-\max(1-V_1, 1-V_3, 0). 
	\label{eq:V_total}
\end{equation} 
Using the expressions of $V$ (Eq.~(\ref{eq:2reflvign})) and $V_2$ (Eq.~(\ref{eq:V2_})), we can write the first term of Eq.~(\ref{eq:V_total}) as
\begin{equation}
	\min(V, V_2, 1) = \min\left(\frac{L_2\alpha_2}{L_1\alpha_1}, \frac{\Psi}{\alpha_1}, 1\right).
	\label{eq:V_total1}
\end{equation} 
Using Eqs.~(\ref{eq:V1_}) and~(\ref{eq:V3_}), the second term reads
\begin{equation}
	\max(1-V_1, 1-V_3, 0) = \max\left(\frac{L_1^* (\alpha_1-\Phi)}{L_1\alpha_1}, \frac{L_2^* (\alpha_2-\Sigma)}{L_1\alpha_1},0 \right).
	\label{eq:V_total2}
\end{equation} 
If positive and not larger than 1, the total vignetting factor (Eq.~(\ref{eq:V_total})), can be used to compute the obstructed mirror effective area. An infinitesimal sector of the primary mirror with polar aperture $\Delta\varphi$, as seen from the off-axis source, has a vignetted geometric area $R_0 V_{\mathrm{tot}}L_1\alpha_1\Delta\varphi$, hence the total effective area is 
\begin{equation}
	A_D(\lambda, \theta)=2R_0 \int_0^{\pi}\!\!V_{\mathrm{tot}}L_1\alpha_1 \,r_{\lambda}(\alpha_1)\,r_{\lambda}(\alpha_2) \,\mbox{d}\varphi.	
	\label{eq:area_total}
\end{equation} 
By substituting Eq.~(\ref{eq:V_total}) into Eq.~(\ref{eq:area_total}), it is straightforward to derive the final expression {\it for the off-axis effective area of the obstructed mirror shell}
\begin{equation}
	A_D(\lambda, \theta)=2R_0 \int_0^{\pi}\!\![(L\alpha)_{\mathrm{min}}- Q_{\mathrm{max}}]_{\ge0} \,r_{\lambda}(\alpha_1)\,r_{\lambda}(\alpha_2) \,\mbox{d}\varphi,	
	\label{eq:area_total_fin}
\end{equation} 
where 
\begin{eqnarray}
	(L\alpha)_{\mathrm{min}} & = & \min\left(L_1\alpha_1, L_2\alpha_2, L_1\Psi\right),
	\label{eq:Lamin}\\
	Q_{\mathrm{max}} & = & \max\left[L_1^*(\alpha_1-\Phi), L_2^*(\alpha_2-\Sigma),0 \right] ,
	\label{eq:Qmax}
\end{eqnarray}
provided that $\alpha_1 \geq 0$, $\alpha_2 \geq 0$, as usual. The $[ \,]_{\ge0}$ brackets in Eq.~(\ref{eq:area_total_fin}) mean that the enclosed expression $(L\alpha)_{\mathrm{min}} - Q_{\mathrm{max}}$, if negative, must be set to zero. For this reason, the integration cannot be carried out independently for the two terms. Moreover, because of the presence of $\Psi$ in Eq.~(\ref{eq:Lamin}) and the different values of $\Phi$ and $\Sigma$ in Eq.~(\ref{eq:Qmax}), the computation is asymmetric with respect to the $y$-axis, also when $\delta =0$. Therefore, the integral in Eq.~(\ref{eq:area_total_fin}) is not equivalent to twice the same integral over $[0, \pi/2]$, {\it unless} {\it all} these conditions are fulfilled (as in Sect.~\ref{geometric}): i)~$\delta =0$, ii)~$\Phi \simeq \Sigma$, and iii)~there is no vignetting of the second kind. 

\subsection{Conditions for obstruction-free mirror shells}\label{no_obst}
We now discuss the conditions for the mirror shell to be obstruction-free. It can be seen, from Eqs.~(\ref{eq:Lamin}) and~(\ref{eq:Qmax}), that there is no obstruction if, and only if, for all $\varphi$ 
\begin{equation}
 	\left\{\begin{array}{l}
 	\alpha_1 < \Phi, \\
 	\alpha_2 < \Sigma, \\
 	\min(L_1\alpha_1, L_2\alpha_2) < L_1\Psi.
	\end{array}\right.\label{eq:obsfree}
\end{equation}
If these inequalities are fulfilled, Eq.~(\ref{eq:area_total_fin}) correctly reduces to Eq.~(\ref{eq:Aeff_fin_offaxis}). The first two conditions are simply met if the maximum values of $\alpha_1$ and $\alpha_2$ do not exceed $\Phi$ and $\Sigma$, respectively. The third condition requires that either $\alpha_1 < \Psi$ {\it or} $\alpha_2 < \frac{L_1}{L_2}\Psi$, which in turn are equivalent\footnote{From Eq.~(\ref{eq:noV2_cond}), it might alternatively follow that $\alpha_1 < \frac{L_1}{L_2}\Psi$ or $\alpha_2~\!\!<~\!\!\Psi$. However, if $L_1 < L_2$ this would still imply that $\alpha_1 < \Psi$, whereas if $L_1 > L_2$ we would still have $\alpha_2 < \Psi < \frac{L_1}{L_2}\Psi$. In both cases, Eq.~(\ref{eq:noV2_cond}) remains valid.} to 
\begin{equation}
	\alpha_1 +\alpha_2 < \left(1+\frac{L_1}{L_2}\right)\Psi, 
	\label{eq:noV2_cond}
\end{equation}
i.e., using Eq.~(\ref{eq:anglesum}),
\begin{equation}
	\frac{2\alpha_0L_2}{L_1+L_2} < \Psi,
	\label{eq:noV2_cond2}
\end{equation}
\begin{figure}
	\resizebox{\hsize}{!}{\includegraphics{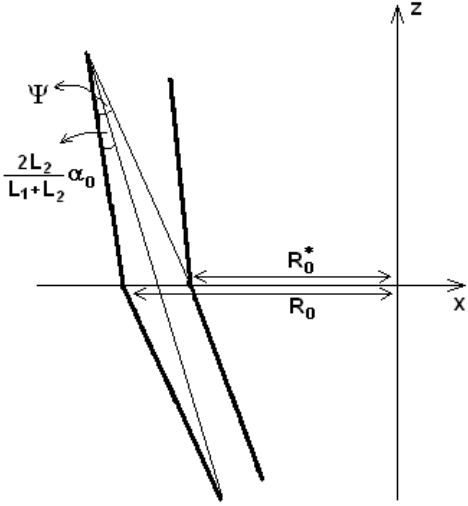}}
	\caption{There is no vignetting at the intersection plane if the distance of the ray reflected at the maximum and the minimum diameter is smaller than the spacing of the two mirror shells.}
	\label{fig:interpr}
\end{figure}
which reduces to $\alpha_0 < \Psi$ if $L_1 = L_2$. The definition of $\Psi$ then allows us to write Eq.~(\ref{eq:noV2_cond2}) as 
\begin{equation}
	\alpha_0\frac{2L_1L_2}{L_1+L_2} <R_0-R_0^*.
	\label{eq:noV2_cond3}
\end{equation}
This result has an immediate geometric interpretation, by noting that left-hand of Eq.~(\ref{eq:noV2_cond3}) is exactly the maximum possible distance from the reflective shell of a ray reflected twice (Fig.~\ref{fig:interpr}). All other rays undergoing a double reflection cannot exceed this distance, which therefore represents the minimum separation for the two shells at $z~=~0$. 

Finally, we define $\tilde{L}$ to be the {\it equivalent length of the mirror}
\begin{equation}
	\frac{2}{\tilde{L}} = \frac{1}{L_1}+\frac{1}{L_2},
	\label{eq:eqlen}
\end{equation}
 and using Eqs.~(\ref{eq:angle1}) and~(\ref{eq:angle2}), we can express the obstruction avoidance conditions as
\begin{eqnarray}
	& &\alpha_0+\delta +\theta < \Phi,\label{eq:noobs1}\\
	& &\alpha_0-\delta +\theta < \Sigma,\label{eq:noobs2}\\
	& &\alpha_0\tilde{L} < R_0-R_0^*. \label{eq:noobs3}
\end{eqnarray}
We note that if $L_1 = L_2 = L$, then also $\tilde{L} = L$, and that the last condition solely depends on the mirror pair geometry, not on the off-axis angle.

\section{Some applications}\label{Application}
\subsection{Analytical expressions for the geometric area}\label{geometric}
As a first application, we derive some expressions for the {\it geometric} area (i.e., in the ideal case $r_{\lambda}(\alpha) =1$ for all $\alpha$) of an obstructed mirror as a function of $\theta$. For simplicity, we only consider the case that $\delta =0$ and $L_1 = L_2 = L_1^* = L_2^*$, and we suppose the mirror to be unobstructed on-axis, i.e., that $\Phi \ge \alpha_0$ (Eq.~(\ref{eq:phi})). Finally, we reasonably assume that $f \gg L$, so that $\Phi \approx \Psi \approx \Sigma$ (Sect.~\ref{obst_param}): consequently, $\Psi \ge \alpha_0$ and there is no vignetting at the intersection plane (as in Fig.~\ref{fig:interpr}). Equation~(\ref{eq:Lamin}) then becomes, as in the unobstructed case,
\begin{equation}
	(L\alpha)_{\mathrm{min}} = L\cdot \min(\alpha_1, \alpha_2).
	\label{eq:La_geom}
\end{equation} 
Adopting $\Phi$ as a unique obstruction parameter, and since $\min(\alpha_1, \alpha_2)+\max(\alpha_1, \alpha_2) = 2\alpha_0$ (Eq.~(\ref{eq:anglesum})), Eq.~(\ref{eq:Qmax}) turns into
\begin{equation}
	Q_{\mathrm{max}} \simeq L\left[2\alpha_0-\min(\alpha_1, \alpha_2)-\Phi\right],
	\label{eq:Qmax_geom}
\end{equation} 
on the condition that both terms are non-negative. Substituting this expression into Eq.~(\ref{eq:area_total_fin}) and using Eq.~(\ref{eq:theta_min}) with $\delta =0$, we obtain
\begin{equation}
	A_D(\theta)=2R_0 L \int_0^{\pi}\!\![\alpha_{\mathrm{min}}- \beta_{\mathrm{max}}]_{\ge0}\,\mbox{d}\varphi,	
	\label{eq:area_geom}
\end{equation}
where we defined 
\begin{eqnarray}
	\alpha_{\mathrm{min}} &=& [\alpha_0-\theta\,|\cos\varphi|]_{\ge0}, \label{eq:alphamin}\\
	\beta_{\mathrm{max}} &=& [\alpha_0+\theta\,|\cos\varphi|-\Phi]_{\ge0} ,\label{eq:betamax}
\end{eqnarray}
and where the $[\,]_{\ge0}$ brackets have the same meaning as those appearing in Eq.~(\ref{eq:area_total_fin}). 

We firstly consider the case $\alpha_0<\Phi < 2\alpha_0$, which implies that $\Phi-\alpha_0 < \alpha_0$. As long as $\theta < \Phi-\alpha_0$, $\alpha_{\mathrm{min}} > 0$ but $\beta_{\mathrm{max}} = 0$ for all $\varphi$, i.e., the mirror is not obstructed and Eq.~(\ref{eq:Ageom_2}) remains valid. 

We now increase $\theta$. Since $\Phi-\alpha_0 < \Phi/2 < \alpha_0$ by hypothesis, we are allowed to consider the case $\Phi-\alpha_0 <\theta< \Phi/2$. For all $\varphi$, we still have $\alpha_{\mathrm{min}}>0$, but this time $\beta_{\mathrm{max}}$ is positive for $\theta\,|\cos\varphi| > \Phi-\alpha_0$: for these values of $\varphi$, the mirror starts to be obstructed. Then Eq.~(\ref{eq:area_geom}) changes into
	\begin{eqnarray}
		A_{\infty}(\theta)&=&4R_0 L \int_{\arccos\frac{\Phi-\alpha_0}{\theta}}^{\pi /2}\!\!\!(\alpha_0-\theta\cos\varphi)\,\mbox{d}\varphi\,\nonumber \\
		&+&4R_0 L\int_0^{\arccos\frac{\Phi-\alpha_0}{\theta}}\!\!\!\! (\Phi-2\theta\cos\varphi)\,\mbox{d}\varphi,
		\label{eq:obs_geom}
	\end{eqnarray}
where the two terms are the unobstructed and the obstructed part of the area, respectively. We note that the two integrands are positive in the respective integration ranges, hence the integration can be carried out and we obtain
\begin{equation}
	\frac{A_{\infty}(\theta)}{A_{\infty}(0)} = 1-\frac{2\theta}{\pi\alpha_0}\left[1+\, S\!\left(\frac{\Phi-\alpha_0}{\theta}\right)\right],
	\label{eq:obs_case1}
\end{equation}
where we normalized the result to $A_{\infty}(0)$, the on-axis geometric area (Eq.~(\ref{eq:Ageom_1})), and defined the non-negative $S(x)$ function,
\begin{equation}
	S(x) = \sqrt{1-x^2}-x\arccos x, \,\, \mbox{with} \,\, 0 \le x \le 1. 
	\label{eq:Sfunction}
\end{equation}
We recognize in the two first terms of Eq.~(\ref{eq:obs_case1}) the usual unobstructed geometric area trend (Eq.~(\ref{eq:Ageom_2})), whereas the term proportional to the $S$ function represents the area lost because of the obstruction. Since $S(1) =0$, Eq.~(\ref{eq:obs_case1}) converges to the unobstructed area trend at $\theta = \Phi-\alpha_0$. 
\begin{figure}
	\resizebox{\hsize}{!}{\includegraphics{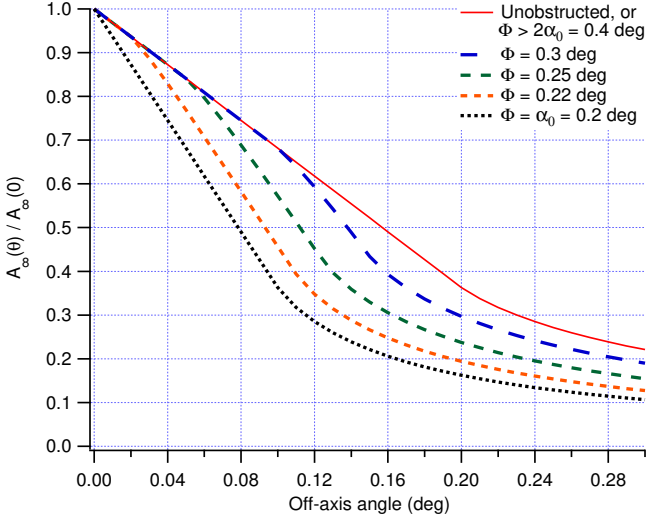}}
	\caption{Normalized geometric area, $A_{\infty}(\theta)/A_{\infty}(0)$ of an obstructed mirror shell with $\alpha_0 = 0.2$~deg, $\delta =0$, and $L_1 = L_2$, as a function of the off-axis angle, for different values of the obstruction parameter $\Phi$, with $\Phi \approx \Psi \approx \Sigma$ and $L = L^*$. The curves are traced using the analytical formulae reported in Sect.~\ref{geometric}. The curve for the unobstructed mirror (the solid line) is also valid for any {\it obstructed} mirror with $\Phi > 2\alpha_0$.}
	\label{fig:geom_area}
\end{figure}

We now consider the case $\theta > \Phi/2$, still for $\Phi<2\alpha_0$ and $\delta =0$. The first term of Eq.~(\ref{eq:obs_geom}) remains unchanged, even if the integration range shrinks because of the larger values of $\theta$. In the second integral, indeed, we have to replace the lower integration limit with $\arccos\!\frac{\Phi}{2\theta}$ to keep the integrand non-negative. This integration limit also guarantees that $\alpha_{\mathrm{min}} > 0$ even for $\theta > \alpha_0$. Because of this change, the resulting expression 
\begin{equation}
	\frac{A_{\infty}(\theta)}{A_{\infty}(0)} = 1-\frac{2\theta}{\pi\alpha_0}\left[1+ \,S\!\left(\frac{\Phi-\alpha_0}{\theta}\right)- 2\, S\!\left(\frac{\Phi}{2\theta}\right)\right],
	\label{eq:obs_case2}
\end{equation}
has an additional term with respect to Eq.~(\ref{eq:obs_case1}).

Finally, we assume that $\Phi \ge 2\alpha_0$. This time $\alpha_0 \le \Phi-\alpha_0$, hence the geometric area decrease deviates from linearity (for $\theta > \alpha_0$) before the mirror begins to be obstructed ($\theta > \Phi-\alpha_0$). The condition $\alpha_{\mathrm{min}} >0$ then implies that the lower integration limit in the first term of Eq.~(\ref{eq:obs_geom}) has to be replaced with $\arccos\frac{\alpha_0}{\theta}$: consequently, $\alpha_{\mathrm{min}} =0$ for $0<\varphi<\arccos\frac{\alpha_0}{\theta}$, and the obstructed term is zero. We conclude that for $\Phi > 2\alpha_0$ and $\delta = 0$ the geometric area trend equals the unobstructed one, i.e., Eq.~(\ref{eq:Ageom_2}) for $\theta < \alpha_0$, and Eq.~(\ref{eq:Ageom_3}) for $\theta > \alpha_0$. As expected, in the limit $\Phi = 2\alpha_0$ Eq.~(\ref{eq:obs_case2}) becomes identical to Eq.~(\ref{eq:Ageom_3}).

As an example, we trace in Fig.~\ref{fig:geom_area} some geometric area curves of a mirror shell for different values of the obstruction parameter $\Phi\ge \alpha_0$, using Eqs.~(\ref{eq:Ageom_2}), (\ref{eq:obs_case1}), and~(\ref{eq:obs_case2}) in the respective intervals of validity. We also plotted the unobstructed geometric area of the mirror (Eqs.~(\ref{eq:Ageom_2}) and~(\ref{eq:Ageom_3})), which is also valid for $\Phi > 2\alpha_0$. Some curves are also validated in Sect.~\ref{Validation} by means of an accurate ray-tracing computation.

\subsection{Design of the mirror module}\label{design}
The obtained results can also be applied to the problem of designing a mirror module. In general, the radius and the length of the outermost shell of the module is assigned on the basis of the allocable space for the optics payload. Starting from this one, shells with decreasing radii are added, leaving a sufficient spacing to minimize the mirror vignetting for off-axis angles within the field of view of the optic. It is then convenient to reduce the obstruction, by increasing $\Phi$ and $\Sigma$ (Sect.~\ref{obst_param}) for every pair of consecutive shells. For any choice of $R_0$, $R_0^*$, and $L_1$, this can be obtained by designing each mirror pair with $L_1^* < L_1$ (Eq.~(\ref{eq:rel1})) and $L_2^*< L_1$ (Eq.~(\ref{eq:rel2})). On the other side, mitigation of the mirror self-vignetting for double reflection (Eq.~(\ref{eq:2reflvign})) requires that $L_2 >L_1$ and $L_2^* >L_1^*$. Hence, if we label the mirror shells with $k$ = 1, 2,\ldots from large to small diameters, a performing module design might consist of mirror shells with 
\begin{equation}
	\cdots \leq L_{1,k+1} \leq L_{2,k+1} \leq L_{1,k} \leq L_{2,k} \leq \cdots,
	\label{eq:lengths}
\end{equation}
i.e., decreasing lengths as the diameter is reduced, and with secondary segments longer than the respective primary {\it but} shorter than the primary of the adjacent mirror shell with larger diameter. A module design with mirror lengths scaled to the diameter has already been studied by Conconi et al.~(\cite{Conconi}) to minimize the defocusing due to the field curvature in the WFXT telescope. 

Therefore, after choosing a sequence of mirror lengths according Eq.~(\ref{eq:lengths}), we apply the obstruction-free conditions (Sect.~\ref{no_obst}). Using Eqs.~(\ref{eq:psi}) and~(\ref{eq:rel2a}), we rewrite Eq.~(\ref{eq:noobs2}) as
\begin{equation}
	\alpha_{0,k}+\theta \leq (R_{0,k}-R_{0,k+1}-\tau_{k+1})\left(\frac{1}{L_{2,k+1}}-\frac{3}{4f}\right),
	\label{eq:cond1}
\end{equation}
where $\tau_{k+1}$ is the thickness of the $(k+1)^{th}$ shell, in general chosen to be proportional to the radius to maintain a constant mirror stiffness throughout the module. If we now define $1/L_{2f, k+1} = 1/L_{2, k+1} - 3/4f$, we can write the last equation as
\begin{equation}
	R_{0, k+1} +\tau_{k+1} \leq R_{0, k}- (\alpha_{0,k}+\theta)L_{2f, k+1}.
	\label{eq:des1}
\end{equation}
In reality, Eq.~(\ref{eq:lengths}) implies that $\Sigma < \Phi$ for all pairs of mirrors, hence Eq.~(\ref{eq:noobs2}) also fulfills Eq.~(\ref{eq:noobs1}). The last relation to be satisfied is Eq.~(\ref{eq:noobs3}), which simply reads
\begin{equation}
	R_{0, k+1} +\tau_{k+1} \leq R_{0, k}- \alpha_{0,k}\tilde{L}_{ k+1}.
	\label{eq:des2}
\end{equation}
We then derive from Eqs.~(\ref{eq:des1}) and~(\ref{eq:des2}) the condition to be fulfilled by the $k^{th}$ couple of shells, in order to avoid obstructions up to an off-axis angle $\theta$
\begin{equation}
	R_{0, k+1} +\tau_{k+1} \leq R_{0, k}- \max\left[\alpha_{0,k}\tilde{L}_{ k+1}, (\alpha_{0,k}+\theta)L_{2f, k+1}\right].
	\label{eq:des}
\end{equation}
If mirror lengths are chosen according to Eq.~(\ref{eq:lengths}), the last relation enables the computation of the maximum possible value of $R_{0,k+1}$ from $R_{0,k}$, $L_{1,k+1}$, $L_{2,k+1}$, and the relation $\tau_{k} = \tau(R_{0, k})$. When applied recursively from the outermost radius inwards, it provides us with the optimal population of mirror shells in the optical module.

\section{Validation with ray-tracing results}\label{Validation}
\begin{figure}
	\resizebox{\hsize}{!}{\includegraphics{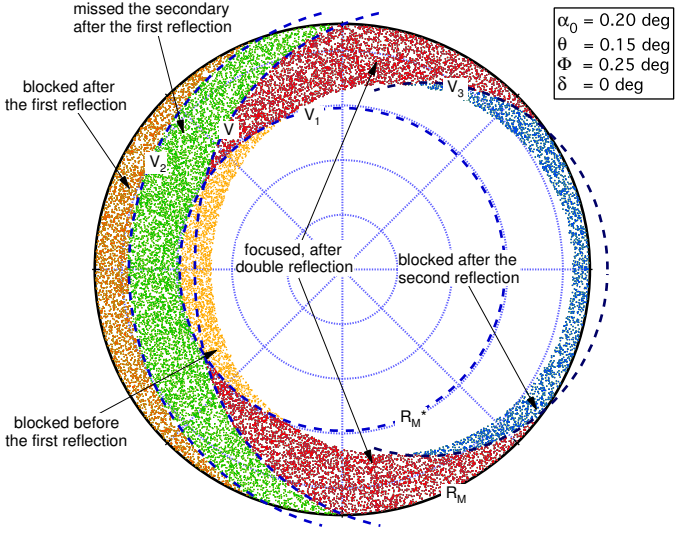}}
	\caption{Initial positions and destinations of 40000 rays at the entrance pupil (points) for a Wolter-I mirror shell with $L = L^*$, and the angular parameters reported in the legend ($\Phi \approx \Psi \approx \Sigma$). The radial scale has been expanded. Only rays that would have struck the primary mirror were traced. The limits of the regions of different vignetting (dashes) are computed from the vignetting coefficients.}
\label{fig:init_pos_trac}
\end{figure}

We hereafter validate the formulae found in the previous sections by comparing them with the results of a detailed ray-tracing. As a first example, we validate the expressions of the vignetting coefficients (Sect.~\ref{obscoeff}). In Fig.~\ref{fig:init_pos_trac}, we show the entrance section of an obstructed Wolter-I mirror shell with $R_0$~=~139.6~mm, $\alpha_0$~=~0.2~deg, $f$~= 10~m, and $L_1 = L_2$~= 300~mm. Rays coming from a source at infinite distance, off-axis by an angle $\theta$~=~0.15~deg, have been traced by simulating their reflection on the mirror: the positions at the entrance pupil of rays that impinged the primary segment are drawn, disregarding the others. The points then fill the geometric section of the primary, as viewed from the direction of the source. The blocking shell has the same focal and length of the reflective one, but a different radius $R_0^*$~=~138.3~mm.

Depending on their initial coordinates, the traced rays can obstructed in various ways, highlighted with different colors in Fig.~\ref{fig:init_pos_trac}, or reach the focal plane if they fall in the red region of the mirror aperture. We note that, in agreement with the discussion in Sect.~\ref{obscoeff}, the obstructions at the entrance (yellow) and the exit pupil (blue) are mainly located close to the intersection plane, i.e., the inner contour of the colored area, while the obstruction after the first reflection (orange), and the vignetting for double reflection (green), mainly occur for rays firstly reflected at locations far from the intersection plane. Moreover, in this case ($\Psi \simeq \Phi > \alpha_0$), the obstruction after the first reflection is ineffective because the orange region, which encloses the rays blocked after the first reflection, is completely surrounded by the green region, which comprises the rays that missed the second reflection. This is in more than qualitative agreement with the analytical tractation: in the same figure we have also traced the analytical expressions of the vignetting coefficients (dashed lines), after translating them into expressions of the radial coordinate (see appendix~\ref{deriV}) and projecting them at the mirror's entrance plane in the direction of the incident rays. We note that the lines perfectly follow the boundaries of the vignetted regions: the agreement shows that the expressions of the vignetting coefficients (Eqs.~(\ref{eq:V1_}) through~(\ref{eq:V3_})) are correct.

\begin{figure}
	\resizebox{\hsize}{!}{\includegraphics{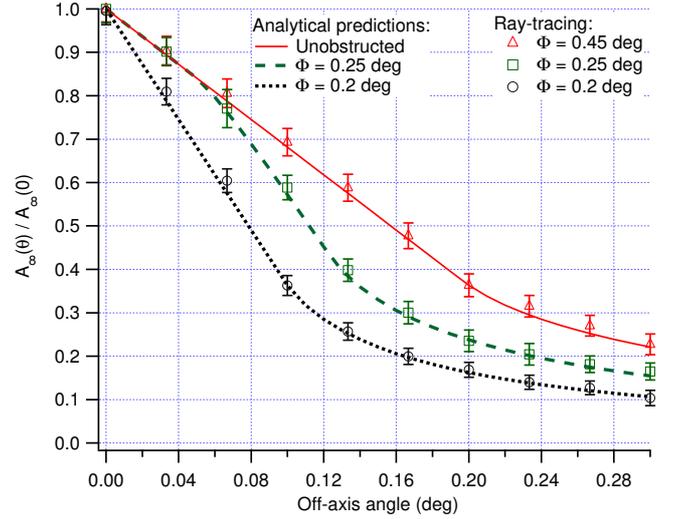}}
	\caption{Comparison between some analytical curves of Fig.~\ref{fig:geom_area} (lines), and the results of an accurate ray-tracing (symbols) for an obstructed Wolter-I mirror with the same $\alpha_0$ = 0.2~deg, $R_0$~=~139.6~mm, $L~=~L^*$~= 300~mm. The outer radius of blocking shell, $R_0^*$, takes on the values 137.24, 138.29, and 138.55~mm, for $\Phi$ to match the values used to draw the analytical curves (but the unobstructed one). The accord between the formulae and the tracing is excellent within the statistical error.}
\label{fig:vign_comp}
\end{figure}

As a second example, we compare the analytical expressions of the geometric area found in Sect.~\ref{geometric} with the findings of the ray-tracing. Figure~\ref{fig:vign_comp} reports some of the normalized geometric area curves (lines) already shown in Fig.~\ref{fig:geom_area}, i.e., for no obstruction and for two values of $\Phi$ in the interval $[\alpha_0, 2 \alpha_0)$. We have also plotted as symbols the geometric area values obtained by applying a ray-tracing routine for a Wolter-I, reflective mirror shell and a co-focal blocking shell with the same length but variable values of $R_0^* < R_0$. Two values of $R_0^*$ (138.55~and 138.29~mm) were chosen to match the two values of $\Phi$ (0.2~and 0.25~deg) we used to trace the analytical curves, while the smallest one (137.24~mm) was selected to return $\Phi$~= 0.45~deg, i.e., a value larger than $2\alpha_0$. 

Inspection of Fig.~\ref{fig:vign_comp} shows that the findings of the two methods are in excellent agreement, within the error bars of the ray-tracing. On the other hand, the ray-tracing routine used is a quite complex code, and the relative computation required a few hours time to reach a few percent accuracy, whilst the analytical curves can be traced immediately and without being affected by statistical errors. We also note that the results of the ray-tracing for $\Phi$~=~0.45~deg are perfectly reproduced by the unobstructed mirror analytical curve, in agreement with the conclusion in Sect.~\ref{geometric} that there is no obstruction at {\it any} off-axis angle if $\Phi > 2\alpha_0$. This happens because, with such a loose mirror nesting, $\theta$ must be larger than $\alpha_0$ for the mirror to start being obstructed. In these conditions, the obstruction is ineffective because it is completely superseded at all polar angles by the vignetting for double reflection. 

\begin{figure}
	\resizebox{\hsize}{!}{\includegraphics{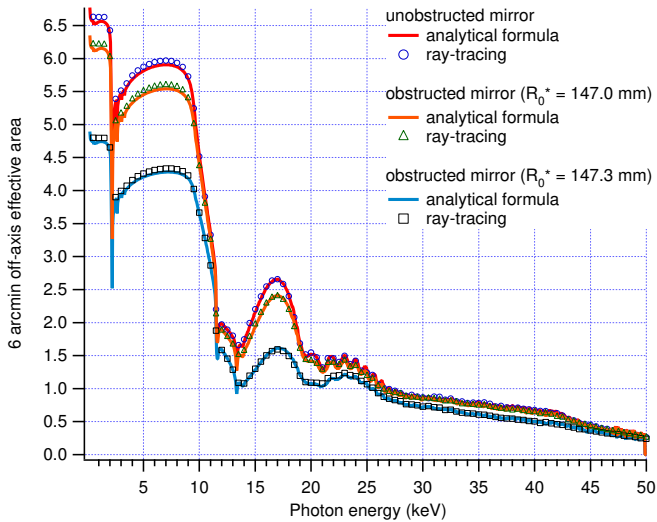}}
	\caption{Effective area of an obstructed mirror shell with $R_0$ = 148.5~mm, $f$~= 10~m, $\alpha_0$ = 0.212~deg, $L$~=$L^*$ =300~mm, and the X-ray source at infinite distance, 6 arcmin off-axis. The curves are computed for three different values of $R_0^*$. The reflective shell has a Pt/C graded multilayer, whose d-spacing variation in depth is a power law, $d(k) = a\,(b+k)^{-c}$, where $k$~=~1 $\ldots$ 200, $a$ = 115.5~\AA,~$b$ =~0.9, $c$ =~0.27. We moreover assumed a Pt thickness-to-d-spacing ratio $\Gamma$ =~0.35, a roughness rms $\sigma$~= 4~\AA, and a stack that ends with Pt on top.}
	\label{fig:obst_sim}
\end{figure}

Finally, we directly compare in Fig.~\ref{fig:obst_sim} some obstructed effective area off-axis curves, as a function of the X-ray energy, as computed from a ray-tracing and using Eq.~(\ref{eq:area_total_fin}). The reflective mirror shell has a fixed radius and a graded multilayer coating (see e.g., Joensen~\cite{Joensen}) to enhance its reflectivity in hard X-rays up to 50 keV and beyond. The characteristics of the reflective shell and the description of the multilayer stack are reported in the figure caption. The off-axis angle $\theta$~=~6~arcmin is the same for all curves, while the outer radius of the blocking shell, $R_0^*$, has been varied. As expected, the effective area decreases as the tightness of the nesting increases (i.e., as $R_0^*$ approaches $R_0$), and moreover the findings of the ray-tracing (symbols) are in perfect agreement with those of the analytical computation (lines). 

We note that the high-energy part of the effective area is less affected by obstruction effects, because it results from reflection at polar angles $\varphi\approx \pm \pi/2$, where $\max(\alpha_1, \alpha_2)$ takes on the smallest value, i.e., $\alpha_0$, and the reflectivity is higher: this is also the angular region where the obstruction is lower (see Fig.~\ref{fig:init_pos_trac}).

\section{Conclusions}\label{Conclusions}
We have developed the analytical formalism for the off-axis effective area of Wolter-I mirror shells, in double cone approximation, which we began to describe in a previous paper (Spiga~et~al.~\cite{Spiga2009}).

We have shown that the analytical expression of the effective area off-axis can be inverted to derive the product of the reflectivity of the two segments (Sect.~\ref{invcomp}). This might be useful to future developments for computing a suitable multilayer recipe to return the desired effective area trend off-axis.

We have found analytical expressions for the vignetting coefficients (Sect.~\ref{obscoeff}), for the three possible sources of obstruction in nested mirror modules, as a function of the azimuthal coordinate of the mirror surface.

Using the vignetting coefficients, we have derived an integral formula (Eq.~(\ref{eq:area_total_fin})) for the obstructed effective area of a Wolter-I X-ray mirror in double cone approximation, with any reflective coating, including multilayers. The computation only requires the standard routines for the reflectivity of the coating, and an integration over the azimuthal coordinate of the mirror shell.

We have obtained analytical expressions of the obstructed geometric area (Sect.~\ref{geometric}) for the case of a source at infinite distance, and applied the result to the problem of designing an optical module that does not suffer from the mutual obstructions of mirrors (Sect.~\ref{design}).

Finally, the results have been validated by means of a comparison with the findings of a detailed ray-tracing (Sect.~\ref{Validation}).

As a final application, we note that each vignetting coefficient can be adapted to estimate the unwanted vignetting caused by collimators aimed at reducing the {\it stray light} in mirror modules (see, e.g., Cusumano~et~al.~\cite{Cusumano}). For example, $V_1$ would quantify the vignetting of the baffle at the entrance pupil, if $R^*_{\mathrm M}$ is interpreted as the outer radius of the collimator ring and $L_1^*$ as its distance from the intersection plane. With analogous substitutions, $V_3$ would represent the vignetting of a baffle located at the exit pupil, and $V_2$ would express the vignetting of the baffle at the intersection plane, even though this kind of baffle can be designed to avoid any obstruction of focused rays (Sect.~\ref{no_obst}). 

\begin{acknowledgements}
This research is funded by ASI (Italian Space Agency, contract I/069/09/0). V.~Cotroneo (Harvard-Smithsonian CfA, Boston, USA) is acknowledged for useful discussions and paper editing. 
\end{acknowledgements}

\appendix

\section{Inversion of the effective area integral}\label{inversion}
We hereafter report the derivation of the inverse integral equation (Eq.~(\ref{eq:refl_prod})), which allows us to derive the reflectivity product from the off-axis effective area (Sect.~\ref{invcomp}). We start from the expression of the unobstructed effective area for $L_1 = L_2$, $\delta =0$, and $0 < \theta \leq \alpha_0$ (Eq.~(\ref{eq:Aeff_alpha})), which we rewrite in terms of the adimensional ratio $E_{\lambda}(\theta)$ (Eq.~(\ref{eq:EA_norm})), 
\begin{equation}
	E_{\lambda}(\theta) =\frac{2}{\pi \alpha_0} \int_{\alpha_0-\theta}^{\alpha_0}\!\frac{\alpha_1\,r_{\lambda}(\alpha_1)\,r_{\lambda}(\alpha_2)}{\sqrt{\theta^2-(\alpha_0-\alpha_1)^2}}\, \mbox{d}\alpha_1,
	\label{eq:Aeff_alpha1}
\end{equation}
where $0 < \theta \leq \alpha_0$. By setting the positive variable $\omega = \alpha_0-\alpha_1$, the integral becomes 
\begin{equation}
	E_{\lambda}(\theta) =\frac{2}{\pi\alpha_0} \int^{\,\theta}_{0}\!\frac{(\alpha_0-\omega)\, T_{\lambda}(\omega)}{\sqrt{\theta^2-\omega^2}}\, \mbox{d}\omega.
	\label{eq:Aeff_omega}
\end{equation}
We now set $K_{\lambda}(\omega)= \omega^2(\alpha_0-\omega) T_{\lambda}(\omega)$, $u = 1/\omega$ and $w = 1/\theta$. Equation~(\ref{eq:Aeff_omega}) can be rewritten as
\begin{equation}
	2\pi\alpha_0\frac{E_{\lambda}(w)}{w} = 4 \int_{w}^{+\infty}\!\frac{u K_{\lambda}(u)}{\sqrt{u^2-w^2}}\, \mbox{d}u.
	\label{eq:Aeff_omega2}
\end{equation}
The right-hand side of Eq.~(\ref{eq:Aeff_omega2}) is the well-known Abel integral. It can be thereby solved for $K_{\lambda}$ (see e.g., Stover~\cite{Stover})
\begin{equation}
	K_{\lambda}(u) = -\,\alpha_0\int_{u}^{+\infty}\!\!\frac{1}{\sqrt{w^2-u^2}}\,\frac{\mbox{d}}{\mbox{d}w}\!\left(\frac{E_{\lambda}(w)}{w}\right)\,\mbox{d}w.
	\label{eq:Aeff_omega3}
\end{equation}
Then, restoring the $T$ function and the $\theta$ and $\omega$ variables, we obtain
\begin{equation}
	T_{\lambda}(\omega) = \frac{\alpha_0}{\omega(\alpha_0-\omega)}\int_{0}^{\omega}\!\!\frac{\theta}{\sqrt{\omega^2-\theta^2}}\,\frac{\mbox{d}}{\mbox{d}\theta}[\theta \cdot E_{\lambda}(\theta)]\:\mbox{d}\theta.
	\label{eq:Aeff_omega4}
\end{equation}
Finally, we substitute $\omega = \alpha_0-\alpha_1$ in Eq.~(\ref{eq:Aeff_omega4}) with $0 \le \alpha_1<\alpha_0$, and change the integration variable from $\theta$ to $t$, where $\theta~=~(\alpha_0-\alpha_1)\sin t$. Hence, we obtain the final result
\begin{equation}
	T_{\lambda}(\alpha_1) = \frac{\alpha_0}{\alpha_1}\int_{0}^{\pi/2}\!\!\sin t \,\left[\frac{\mbox{d}}{\mbox{d}\theta}(\theta \cdot E_{\lambda}(\theta))\right]_{\theta = \theta(t)}\!\!\!\!\!\mbox{d}t,
	\label{eq:refl_product}
\end{equation}
where the [ ] brackets mean that the enclosed expression has to be evaluated at $(\alpha_0-\alpha_1)\sin t$. We have so obtained Eq.~(\ref{eq:refl_prod}).

\section{Derivation of the vignetting coefficients}\label{deriV}
We hereafter report the detailed derivation of the vignetting coefficients (Sect.~\ref{obscoeff}). The geometry of mirror shell obstructions and the meaning of symbols are explained in Figs.~\ref{fig:wolter1}, \ref{fig:obstruct}, and~\ref{fig:obstruct_param}.

\subsection{Vignetting at the entrance pupil, $V_1$}\label{V1}
This obstruction is caused by the shadow cast by the primary segment of the blocking shell onto the primary segment of the reflective shell. For this to occur, the shadow of the edge of the blocking shell at ${\underline r}_0 = (R_{\mathrm M}^*\cos\varphi_0, R_{\mathrm M}^*\sin\varphi_0, L_1^*)$ has to intersect the reflective shell at some $\underline{r}_1 = (r_1\cos\varphi_1, r_1\sin\varphi_1, Z_1)$, with $0<Z_1<L_1$. The shaded part of the mirror length is then $Z_1/L_1$.

To find $Z_1(\varphi_1)$, we consider the generic ray emerging from the source, which is located at ${\underline S} = (D\sin\theta, 0, D\cos\theta)$, and passing by ${\underline r}_0$. Now, ${\underline r}_1$, the ray intersection point on the reflecting primary, whose equation is
\begin{eqnarray}
	z_1 = \frac{r_1-R_0}{\alpha_0}, & \mbox{with}& r_1>R_0,
	\label{eq:cone1}
\end{eqnarray}
in conical approximation, is identified via the vector equation
\begin{equation}
	({\underline S}-{\underline r}_0)\times({\underline r}_1-{\underline r}_0) =0, 
	\label{eq:V1_eq}
\end{equation}
where $\times$ denotes a cross product. This returns two independent scalar equations
\begin{equation}
\left\{\begin{array}{l}
	\left(1-\frac{L_1^*}{D}\right)r_1\cos\varphi_1 + \frac{\delta^*\cos\varphi_0-\theta}{\alpha_0}(r_1-R_0) = R_{\mathrm M}^*\cos\varphi_0-L_1^*\theta\\
	\left(1-\frac{L_1^*}{D}\right)r_1\sin\varphi_1 +\frac{\delta^*\sin\varphi_0}{\alpha_0}(r_1-R_0) = R_{\mathrm M}^*\sin\varphi_0
\end{array}\right.\label{eq:V1_eqs},
\end{equation}
where we defined $\delta^* = R_{\mathrm M}^*/D$ to be the beam divergence at the blocking mirror shell. 

The impact position is obtained by solving the previous equations for $r_1$ and $\varphi_1$. If $\theta =0$, the solution is expected to be independent of $\varphi_0$, and the first points to be shaded are near the intersection plane, i.e., $r_1 \approx R_0$. To treat the general case of $\theta \ge 0$, we search for a perturbative solution in the approximation of small $\theta$, so we set $r_1 = R_0 + \xi$ and $\varphi_1 = \varphi_0 +\varepsilon$, with $0 < \xi \ll R_0$ and $|\varepsilon| \ll \varphi_0$. By substituting in Eqs.~(\ref{eq:V1_eqs}), approximating to the first order, and solving, we obtain a linear system whose solutions are
\begin{eqnarray}
	\xi & \simeq & -\frac{\Phi-(\alpha_0+\delta-\theta\cos\varphi_0)}{\alpha_0+\delta^*-\theta\cos\varphi_0}L_1^*\alpha_0\label{eq:xi0},\\
	\varepsilon& \simeq&\frac{L_1^*}{R_0}\,\frac{\delta^*+\Phi}{(\alpha_0+\delta^*-\theta\cos\varphi_0)}\,\theta\sin\varphi_0\,\label{eq:eps0},
\end{eqnarray}
where we have used the definition of $\Phi$ (Eq.~(\ref{eq:phi})) and neglected terms proportional to $\alpha_0\delta$.

We note that $\varepsilon$ is of the order of $\theta$ or less, i.e., it is negligible with respect to $\varphi_0$ itself. We can then assume that $\varphi_1 \approx \varphi_0$. Moreover, we know from Sect.~\ref{review} also that $\varphi_1 \approx \varphi_2$, thus we denote with $\varphi$ the nearly-common value of all these polar angles. We then rewrite Eq.~(\ref{eq:xi0}), using Eq.~(\ref{eq:cone1}), as 
\begin{equation}
	Z_1(\varphi) \simeq -L_1^*\frac{\Phi-(\alpha_0+\delta-\theta\cos\varphi)}{\alpha_0+\delta^*-\theta\cos\varphi}.
	\label{eq:z1_0}
\end{equation}
Clearly, $Z_1>0$ only if $\xi >0$. If the spacing of the two mirrors is small enough, we may also approximate $\delta^* \approx \delta$. The first vignetting factor is thereby $V_1 = 1-\frac{Z_1}{L_1}$, i.e., recalling the definition of $\alpha_1$ (Eq.~(\ref{eq:angle1})),
\begin{equation}
	V_1(\alpha_1) \simeq 1+ \frac{L_1^*(\Phi-\alpha_1)}{L_1\alpha_1},
	\label{eq:V1}
\end{equation}
if positive and less than 1. We have so obtained Eq.~(\ref{eq:V1_}). As usual, if this expression returns a negative value at some $\varphi'$, then $V_1(\varphi') = 0$, or, if larger than one, $V_1(\varphi') = 1$. If $L_1 = L_1^*$ we find that $V_1$ takes the simple form of 
\begin{equation}
	V_1(\alpha_1) \simeq \frac{\Phi}{\alpha_1},
	\label{eq:V1_altfin}
\end{equation}
where in this case we also note that, {\it if} $V_1$ were positive and smaller than one for all $\varphi$, the geometric area of the obstructed {\it primary} segment would become 
\begin{equation}
	A_1(\theta) = 2R_0L_1\int_0^{\pi}\!\alpha_1V_1(\alpha_1)\,\mbox{d}\varphi,
	\label{eq:A1g_obs}
\end{equation}
which -- as expected -- returns the area of the corona delimited by the largest radii of the two shells 
\begin{equation}
	A_1(\theta) = 2\pi R_0L_1\Phi =2\pi R_0(R_{\mathrm M}-R_{\mathrm M}^*),
	\label{eq:A1g_obs2}
\end{equation}
where we have used the relation $R_{\mathrm M} \simeq R_0+\alpha_0L_1$. 

\subsection{Vignetting at the intersection plane, $V_2$}\label{V2}
In this case, vignetting occurs after the first reflection from the primary segment of the inner shell, i.e., a point of the blocking shell's outer surface at $z=0$, with coordinates ${\underline r}_0 = (R_0^*\cos\varphi_0, R_0^*\sin\varphi_0, 0)$, may intercept a ray reflected by the primary segment of the reflective shell at $\underline{r}_1=(r_1\cos\varphi_1, r_1\sin\varphi_1, Z_1)$, with $0< Z_1<L_1$. If it does, obstruction occurs at $Z_1(\varphi_1) < z < L_1$. The coordinate $Z_1$ is located via the vector equation
\begin{equation}
	{\underline k}_1\times({\underline r}_1-{\underline r}_0) =0,
	\label{eq:V2_eq}
\end{equation}
where $\underline{k}_1$ is the ray direction after the first reflection. This unit vector has the following expression\footnote{From the expression of the $\underline{k}_1$ vector, it is easy to derive the {\it stray light} pattern on a detector at a generic distance $d$, i.e., at $z = -d$. The final coordinates of rays generated close to the intersection plane are
\begin{eqnarray}
x(\varphi) &=& R_0\cos\varphi-\frac{k_{1,x}}{k_{1,z}}d\nonumber,\\
y(\varphi) &=& R_0\sin\varphi-\frac{k_{1,y}}{k_{1,z}}d\nonumber,
\end{eqnarray}
and, substituting the expression of the $\underline{k}_1$ vector components, we obtain the parametric equation of the pattern
\begin{eqnarray}
x(\varphi) &=& [R_0-(2\alpha_0+\delta)d]\cos\varphi+\theta d \cos2\varphi,\nonumber\\
y(\varphi) &=& [R_0-(2\alpha_0+\delta)d]\sin\varphi+\theta d \sin2\varphi.\nonumber
\end{eqnarray}} (Spiga~et al.~\cite{Spiga2009})
\begin{equation}
	 \underline{k}_1 \simeq \left(
	\begin{array}{c}
	-\theta+(\delta-2\alpha_1)\cos\varphi_1\\
	(\delta-2\alpha_1)\sin\varphi_1\\
	-1
	\end{array}\right).
\label{eq:exit_dir1}
\end{equation}
After some manipulation of Eq.~(\ref{eq:V2_eq}), we obtain the two scalar equations
\begin{equation}
\left\{\begin{array}{ccc}
	(\alpha_0\cos\varphi_1+k_{1x})r_1 &=& k_{1x}R_0+\alpha_0R_0^*\cos\varphi_0 \\
	(\alpha_0\sin\varphi_1+k_{1y})r_1 &=& k_{1y}R_0+\alpha_0R_0^*\sin\varphi_0
	\end{array}\right.\label{eq:V2_eqs},
\end{equation}
and, using Eq.~(\ref{eq:exit_dir1}), we find that the solution of Eqs.~(\ref{eq:V2_eqs}) for $\theta =0$ is $\varphi_0 = \varphi_1$ and 
\begin{equation}
	r_1^{(0)}=R_0+ \frac{\alpha_0}{\alpha_0+\delta}(R_0-R_0^*).
	\label{eq:soluz0}
\end{equation}
Since $R_0 > R_0^*$, we have $r_1^{(0)} > R_0$, hence $Z_1^{(0)} > 0$ (Eq.~(\ref{eq:cone1})). If $\theta > 0$, we set $r_1 = r_1^{(0)}+\xi$, $\varphi_0 = \varphi_1+\varepsilon$, and proceed as in Sect.~\ref{V1}. The solution to a first order approximation is
\begin{equation}
	\xi \simeq \frac{\alpha_0(\alpha_0+\delta-\alpha_1)}{\alpha_1(\alpha_0+\delta)}(R_0-R_0^*).
	\label{eq:xi1}
\end{equation}
We are not interested in the exact expression of $\varepsilon$, which is of the order of $\theta$, so we can assume again that $\varphi_0 \approx \varphi_1$ and neglect the $\varphi$'s subscript. In contrast, from $\xi$ and Eq.~(\ref{eq:soluz0}) we can derive an expression for $r_1$, and using Eq.~(\ref{eq:cone1}) we obtain $Z_1$
\begin{equation}
	Z_1(\varphi) \simeq \frac{R_0-R_0^*}{\alpha_1},
	\label{eq:z1_1}
\end{equation}
which is always {\it non-negative}. All points with $z > Z_1$ are then obstructed. The resulting vignetting coefficient is $V_2 = Z_1/L_1$, i.e., using the definition of $\Psi$ (Eq.~(\ref{eq:psi})),
\begin{equation}
	V_2(\alpha_1) \simeq \frac{\Psi}{\alpha_1}.
	\label{eq:V2}
\end{equation}
We have so obtained Eq.~(\ref{eq:V2_}).

\subsection{Vignetting at the exit pupil, $V_3$}\label{V3}
In this case, the obscuration occurs after the second reflection, on the secondary segment of the blocking shell. A generic point of the blocking shell at $z = -L_2^*$, i.e., ${\underline r}_0 = (R_{\mathrm m}^*\cos\varphi_0, R_{\mathrm m}^*\sin\varphi_0, -L_2^*)$, may intercept a ray after it was reflected at $\underline{r}_2 = (r_2\cos\varphi_2, r_2\sin\varphi_2, Z_2)$, with $-L_2<Z_2<0$. If it does, the point $\underline{r}_2$ is located along with the usual vector equation
\begin{equation}
	{\underline k}_2\times({\underline r}_2-{\underline r}_0) =0,
	\label{eq:V3_eq}
\end{equation}
where the z-coordinate of the secondary segment is given by
\begin{eqnarray}
	z_2 = \frac{r_2-R_0}{3\alpha_0}, & \mbox{with}& r_2<R_0,
	\label{eq:cone2}
\end{eqnarray}
and $\underline{k}_2$ is the direction of the ray after the second reflection,
\begin{equation}
	\underline{k}_2 =\underline{k}_1 -2(\underline{k}_1\cdot \underline{n}_2)\, \underline{n}_2,
	\label{eq:k2_def}
\end{equation}
because the tangential component is conserved, while the normal component to the surface reverses its sign in the reflection. To the small angle approximation, since $\varphi_2 \simeq \varphi_1$ (Sect.~\ref{review}), 
\begin{equation}
	\underline{n}_2 \approx \left(
	\begin{array}{c}
		-\cos{\varphi_1}\\
		-\sin{\varphi_1}\\
		3\alpha_0
	\end{array}
	\right),
	\label{eq:n2}
\end{equation}
and $\underline{k}_1$ is given by Eq.~(\ref{eq:exit_dir1}). The final ray direction\footnote{From Eq.~(\ref{eq:exit_dir2}) we obtain that all rays reflected twice at locations close to the intersection plane ($r = R_0$) converge to a single point $F~=~(-\theta f', 0, -f')$, as expected, with
\begin{displaymath}
	f' = \frac{R_0}{4\alpha_0-\delta},
\end{displaymath}
which, recalling Eq.~(\ref{eq:Rf}) and that $\delta = R_0/D$, becomes the usual {\it conjugate points formula},
\begin{displaymath}
	\frac{1}{f} = \frac{1}{f'}+\frac{1}{D}.
\end{displaymath}
Focusing at that point, indeed, does not exactly occur for points more distant from the intersection plane. This is a well-known limit of the double cone approximation, even if it does not affect our computation of the effective area within large limits (Sect.~\ref{review}).} is then
\begin{equation}
	\underline{k}_2 \simeq \left(
		\begin{array}{c}
		-\theta+(\delta-4\alpha_0)\cos\varphi_1\\
		(\delta-4\alpha_0)\sin\varphi_1\\
		-1
		\end{array}
	\right).
	\label{eq:exit_dir2}
\end{equation}
Setting $\varphi_2 = \varphi_1$ in Eq.~(\ref{eq:V3_eq}), we obtain the two independent equations
\begin{equation}
	\left\{\begin{array}{ccc}
		(3\alpha_0\cos\varphi_1+k_{2x})r_2 &=& k_{2x}R_{\mathrm m}'+3\alpha_0R_{\mathrm m}^*\cos\varphi_0\\
		(3\alpha_0\sin\varphi_1+k_{2y})r_2 &=& k_{2y}R_{\mathrm m}'+3\alpha_0R_{\mathrm m}^*\sin\varphi_0
	\end{array}\right.\label{eq:V3_eqs},
\end{equation}
where we set $R_{\mathrm m}' \simeq R_0-3\alpha_0L_2^*$. The solution for $\theta =0$ is $\varphi_0 = \varphi_1$ and 
\begin{equation}
	r_2^{(0)}= R_{\mathrm m}'+\frac{3\alpha_0}{\alpha_0-\delta}(R_{\mathrm m}'-R_{\mathrm m}^*).
	\label{eq:soluz0_2}
\end{equation}
If $\theta >0$, we set $r_2 = r_2^{(0)}+\xi$, $\varphi_0 = \varphi_1+\varepsilon$ as in Sects.~\ref{V1} and~\ref{V2}. Substituting this into Eq.~(\ref{eq:V3_eqs}), approximating at the first order and solving the linear system, we obtain $\varphi_0 \approx \varphi_1$ and the expression of $\xi$. Adding this to $r_2^{(0)}$ we obtain
\begin{equation}
	r_2= R_{\mathrm m}'+\frac{3\alpha_0}{\alpha_2}(R_{\mathrm m}'-R_{\mathrm m}^*),
	\label{eq:soluz1_2}
\end{equation} 
where $\alpha_2$ is given by Eq.~(\ref{eq:angle2}). The secondary mirror region at $\varphi_1$ with radial coordinate between $r_2$ and $R_0$ is then obstructed: this is equivalent to an obscuration on the primary mirror between $z =0$ and some $z = Z_1 < L_1$ (Fig.~\ref{fig:obstruct}). The correspondence between $r_1$ and $r_2$ is given by the equation (Spiga~et~al.~\cite{Spiga2009}) 
\begin{equation}
	r_2 \simeq \frac{2(\alpha_0+\alpha_1)R_0-3\alpha_1r_1}{\alpha_2}.
	\label{eq:r2}
\end{equation}
By comparing Eqs.~(\ref{eq:soluz1_2}) and~(\ref{eq:r2}), we find $r_1$, and the corresponding value of $Z_1$ using Eq.~(\ref{eq:cone1}),
\begin{equation}
	Z_1 \simeq \frac{(\alpha_2+3\alpha_0)L_2^*-(R_0-R_{\mathrm m}^*)}{\alpha_1},
	\label{eq:z1_2}
\end{equation}
from which, recalling that $\alpha_2 = 2\alpha_0 -\alpha_1$ (Eq.~(\ref{eq:anglesum})), we obtain the vignetting coefficient, $V_3 = 1-Z_1/L_1$,
\begin{equation}
	V_3(\alpha_1) \simeq 1+\frac{L_2^*(\Sigma-\alpha_2)}{L_1\alpha_1},
	\label{eq:V3}
\end{equation}
where we have used the definition of $\Sigma$ (Eq.~(\ref{eq:sigma})). We have thereby found Eq.~(\ref{eq:V3_}).

\end{document}